\documentclass[conference]{IEEEtran}

\usepackage{graphicx}
\usepackage{xcolor}
\usepackage{cite}
\usepackage{multirow}
\usepackage{rotate}
\usepackage{subfigure}
\usepackage{array}
\usepackage{subfig}
\usepackage{url}

\DeclareGraphicsExtensions{.pdf,.png,.jpg}
\graphicspath{{figures/}}

\ifCLASSINFOpdf
\else
\fi
\begin{document}

\title{Are You Ready? Towards the Engineering of Forensic-Ready Systems}


\author{\IEEEauthorblockN{George Grispos$^{1}$, Jes\'{u}s Garc\'{i}a-Gal\'{a}n$^{1}$, Liliana Pasquale$^{2}$, Bashar Nuseibeh$^{1,3}$} \\
	\IEEEauthorblockA{$^{1}$Lero -- The Irish Software Research Centre, University of Limerick, Ireland \\
		$^{2}$Lero -- The Irish Software Research Centre, University College Dublin, Ireland \\ 
		$^{3}$The Open University, Milton Keynes, United Kingdom \\ 
		george.grispos@lero.ie, jesus.galan@lero.ie, liliana.pasquale@lero.ie, bashar.nuseibeh@lero.ie}
}

\maketitle

\begin{abstract}\footnote{\textbf{Recommended Citation}: \textit{G. Grispos, J. Garc\'{i}a-Gal\'{a}n, L. Pasquale and B. Nuseibeh (2017). Are You Ready? Towards the Engineering of Forensic-Ready Systems. IEEE 11th International Conference on Research Challenges in Information Science, Brighton, United Kingdom.}}. As security incidents continue to impact organisations, there is a growing demand for systems to be `forensic-ready' -- to maximise the potential use of evidence whilst minimising the costs of an investigation. Researchers have supported organisational forensic readiness efforts by proposing the use of policies and processes, aligning systems with forensics objectives and training employees. However, recent work has also proposed an alternative strategy for implementing forensic readiness called \textit{forensic-by-design}. This is an approach that involves integrating requirements for forensics into relevant phases of the systems development lifecycle with the aim of engineering forensic-ready systems. While this alternative forensic readiness strategy has been discussed in the literature, no previous research has examined the extent to which organisations actually use this approach for implementing forensic readiness. Hence, we investigate the extent to which organisations consider requirements for forensics during systems development. We first assessed existing research to identify the various perspectives of implementing forensic readiness, and then undertook an online survey to investigate the consideration of requirements for forensics during systems development lifecycles. Our findings provide an initial assessment of the extent to which requirements for forensics are considered within organisations. We then use our findings, coupled with the literature, to identify a number of research challenges regarding the engineering of forensic-ready systems. 
\end{abstract}


%
\IEEEpeerreviewmaketitle


\section{Introduction}

According to a recent industrial survey, a quarter of all businesses in the United Kingdom detected a security incident in the previous twelve months \cite{BreachesSurvey2016}. The consequences of such incidents for an organisation can include significant financial losses, a loss of customer confidence and a reduction in business reputation \cite{Ponemon15Cost}. When a security incident occurs, organisations usually respond by conducting a forensic investigation \cite{grisposrethinking}. The purpose of this investigation is to collect and analyse forensic data such as log files and network traffic, in order establish the cause of the incident and how it can be prevented in the future \cite{casey2011digital}.

In order to examine the causes of an incident, investigators rely on the availability of forensic data from affected systems~\cite{carrier2003getting}. However, such data might not always be available for a variety of reasons including limited physical access, short data retention times and the costs associated with conducting such investigations \cite{grispos2015security,stephenson2003conducting}. As a result, investigators may not be able to identify the causes of an incident \cite{stephenson2003conducting}. Hence, there have been increasing calls for organisations to maximise their potential use of evidence and minimise the costs of an investigation by ensuring that their systems and infrastructure are `forensic-ready' \cite{rowlingson2004ten,tan2001forensic}. Researchers have supported forensic readiness efforts by proposing that organisations implement policies and processes \cite{rowlingson2004ten}, align systems with forensics objectives \cite{reddy2013architecture} and the correct training of employees \cite{rowlingson2004ten}.

In the past few years, researchers have proposed an alternative forensic readiness strategy called \textit{forensic-by-design}~\cite{ ab2016forensic,mink2016next}. Conceptually, forensic-by-design is similar to security-by-design, where requirements for forensics are integrated into relevant phases of the systems development lifecycle, with the objective of engineering forensic-ready systems \cite{ab2016forensic}. While this approach has been discussed in the literature, no previous research has examined the extent to which organisations actually implement this approach to achieve forensic readiness.

Hence, our paper provides an initial assessment of the extent to which requirements for forensics are considered within organisations. As a method, we analysed the literature to examine the various perspectives of implementing forensic readiness. We then undertook an online survey to examine the industrial perspective on the `forensic-by-design' approach to forensic readiness. The survey targeted practitioners traditionally involved in the development lifecycle, as well as security and forensics professionals. The survey results, coupled with the findings from the literature are then used to propose research challenges that emerge from our work. 

The paper is structured as follows. Section \ref{sec:method} describes the research methodology and Section \ref{sec:litreview} reviews various perspectives on implementing forensic readiness. Section \ref{sec:Survey} presents the survey results and discusses the threats to the validity of our study. Section \ref{sec:Challenges} presents the research challenges that emerge from our work and Section \ref{sec:Conclusions} draws conclusions.

\section{Research Methodology}
\label{sec:method}

The research described in this paper uses a two-step exploratory research approach \cite{shull2008guide}. We first assessed the literature for existing research on implementing forensic readiness within organisations, as well as relevant literature related to the use of forensic-by-design approaches. After assessing the literature, we then undertook an online survey to investigate the industrial perspective on the use of forensic-by-design approaches. The survey was designed and hosted on LimeService \cite{Lime16URL}, an online survey platform, from March to August 2016. The survey research method was chosen because it provides a low-cost approach for collecting a large cross-section of subjective responses from respondents \cite{Run2012CaseStudy}. The survey consisted of twenty `closed' questions\footnote{\url{http://spare.lero.ie/pdf/survey_questions.pdf}}. However, the number of questions actually presented to the participant was dependent upon the answers provided during the survey. The survey was divided into three parts. In part one participant information was collected, while the second and third parts of the survey presented questions on requirements for investigations and forensic data, respectively. The questions consisted of 3, 5 and 7-point semantic scale answers. Where required, we also included a text-box for respondents to provide textual answers. These type of questions allow for eliciting subjective assessment, while providing some flexibility when interpreting the results \cite{Run2012CaseStudy}. The questions were approved by an Ethics Committee at the University of Limerick. In total, the survey took approximately twelve minutes to complete.  

The survey targeted individuals via LinkedIn groups where the focus of the group was software engineering/development, requirements engineering, security or digital forensics. A short post was placed in each group, inviting the members to participate in the survey. The survey targeted individuals traditionally involved in the development lifecycle, as well as security and forensics professionals. We believe that responses from the later individuals would benefit our study due to their involvement in security incidents and forensic investigations~\cite{grispos2016enhancement,casey2011digital}. This recruitment approach aligns with the purpose and relevance of the study targeting industrial professionals.

After closing the survey, all the data including any respondent comments were aggregated and analysed using a combination of a qualitative and quantitative approach. The idea behind the analysis was to identify trends, patterns and anomalies. Based on the survey findings and the analysis of the literature, we then identified research challenges related to the engineering of forensic-ready systems.

\section{Analysis of Forensic Readiness Approaches}
\label{sec:litreview}	

Digital forensics is often part of an organisation's security incident response capability \cite{freiling2007common}. While the objective of incident response is to restore service and learn about a security incident, digital forensics is concerned with the collection and analysis of forensic data, which can then be used as evidence in court \cite{freiling2007common}. Several approaches have been proposed for conducting forensic investigations (e.g., \cite{freiling2007common,carrier2003getting}). 

Traditionally, investigators only attempt to collect and analyse forensic data concerning an incident after it has occurred~\cite{grispos2016enhancement}. However, Rowlingson \cite{rowlingson2004ten} argues that because organisations ignore what happens to their system(s) prior to an incident, data that is required for an investigation will either exist and is preserved by the system, or it does not exist and the incident cannot be investigated effectively. These concerns have prompted recommendations that organisations take a more proactive stance to digital forensics and structure their environment to retain data required for investigations \cite{tan2001forensic,rowlingson2004ten}. This stance is known as \textit{forensic readiness} \cite{rowlingson2004ten}. Researchers have broadly defined forensic readiness solutions along the lines of implementing policies and processes, aligning systems with forensics objectives and the training of employees.

\subsubsection{\textbf{Policies}}

Policies can provide a cohesive structure by indicating what needs to be done within an organisation~\cite{grisposcloud,harris2010cissp}. As a result, researchers have suggested that organisations implement well--defined policies to support forensic readiness~\cite{rowlingson2004ten,barske2010digital}. Barske, et al. \cite{barske2010digital} assert that these policies include how systems are monitored, under what conditions data is preserved, data retention times, as well as policies on how investigations are to be undertaken. Rowlingson \cite{rowlingson2004ten} adds that policies should also cover topics such as identifying data sources, determining data collection and secure data storage. Endicott-Popovsky, et al. \cite{endicott2007theoretical} argue that such policies should be regularly audited to ensure continuous forensic readiness.

\subsubsection{\textbf{Systems and Frameworks}}

Tan \cite{tan2001forensic} proposed the concept of `system forensic readiness', which consists of identifying what data sources exist within an organisation and what happens to data that could be used as evidence. This concept has been well discussed in the literature \cite{barske2010digital,pooe2012conceptual,rowlingson2004ten,yasinsac2001policies}. Separately, other researchers have suggested using technology to enhance forensic readiness \cite{reddy2013architecture,grobler2007digital}. The idea is that technology can be used to assist investigators to identify malicious users and their activities. For example, Reddy and Venter \cite{reddy2013architecture} proposed an architecture for a forensic readiness management system that includes monitoring for security incidents, escalation rules, storing data, recording training and descriptions of business processes. Similarly, Grobler and Louwrens \cite{grobler2007digital} put forward the idea that organisations implement digital evidence record management systems to retain potential evidence in its original format.

\subsubsection{\textbf{Processes}}

While some researchers have proposed using technology, others have argued that well--defined processes can help enhance forensic readiness \cite{yasinsac2001policies,reddy2009forensic,ab2016forensic, barske2010digital}. For example, supporting forensic investigators with a well-defined investigation process can help reduce mistakes during investigations \cite{yasinsac2001policies,reddy2009forensic}. Moreover, data preservation and collection processes can also increase the speed of an investigation and decrease the cost of reacquiring any data that was not acquired correctly \cite{casey2005case, rowlingson2004ten}. Tan \cite{tan2001forensic} adds that organisations should have processes for defining how and what logging is done, how systems are examined, and how forensic data is acquired during investigations. Other researchers have suggested that organisations have processes to identify incident scenarios and the forensic data required for these scenarios \cite{rowlingson2004ten,grobler2007digital, endicott2007theoretical}.

\subsubsection{\textbf{People}}

Researchers have contended that organisations should establish multidisciplinary forensic readiness teams~\cite{dlamini2014requirements,grispos2015security, pooe2012conceptual,tan2001forensic}. Dalmani, et al. \cite{dlamini2014requirements} argue that adding legal experts to this team can assist investigators with determining the scope of potential evidence to be collected. Grispos, et al. \cite{grispos2015security} argue that integrating a human resources department within this team can assist investigators with employees who are involved in an incident. However, before this forensic readiness team can assist investigators it must receive the correct training to fulfil its objectives \cite{ hoolachan2010organizational,rowlingson2004ten}. For example, investigators need to be trained on how to conduct a forensic investigation correctly. A failure to do so can increase investigation costs and result in compromised evidence \cite{hoolachan2010organizational}.

\subsubsection{\textbf{Forensics-By-Design Approaches}}

In the past few years, researchers have proposed an alternative forensic readiness strategy known as \textit{forensic-by-design} \cite{mink2016next,ab2016forensic}. Ab Rahman, et al. \cite{ab2016forensic} proposed a forensic-by-design framework to integrate forensic tools and best practices in the design and development of cyber-physical cloud systems. Ab Rahman, et al. \cite{ab2016cloud} later expanded their approach to include incident handling capabilities and evaluated their approach using cloud storage services. Mink, et al. \cite{mink2016next} discussed the impact of digital forensics on next generation aircraft systems and concluded that forensic-by-design principles should be integrated into such systems. While researchers have proposed the use of forensic-by-design to enhance forensic readiness, no previous research has examined the extent to which organisations actually use this approach during systems development.

\section{Survey of Practitioners}
\label{sec:Survey}

126 participants attempted our survey. Filtering out incomplete and invalid responses resulted in 94 valid responses (74.6\% completion rate). Nearly 55\% of the respondents indicated that they were involved in software development roles. This includes software/system engineers (27 responses), software/project managers (14 responses) and programmers/developers (10 responses). The remaining respondents included consultants (17 responses), security engineers (5 responses), security managers (3 responses), security analyst (1 response) and other (17 responses). The participants indicated that they were experienced professionals, with 75 (79.8\%) out of the 94 respondents stating that they had more than 10 years experience in an IT-based job role. More than three quarters of the participants came from Europe (57\%) and North America (25\%). The remaining participants included 9\% from Asia, 3\% each from the Middle East and Africa, 2\% from South America and 1\% from Central America. 

The participants represented a diverse set of industries including Information Technology (47\%), Financial Services (12\%), Telecommunications (6\%), Aerospace (5\%), and Government (5\%). We then asked the participants about their experience with requirements management. Nearly 98\% of the participants indicated that they had been involved in the elicitation of software requirements and just over 77\% indicated that they had been involved in the elicitation of security requirements. The following subsections present our survey findings from the perspective of considering requirements for investigations and forensics data, as well as identifying the practitioner's perspective on these types of requirements.

\subsection{Requirements for Investigations}
\label{subsec:investigationreqs} 

Initial questions asked our survey participants if their organisation had in the past elicited requirements concerning the detection, investigation, eradication and recovery of incidents. These are investigation phases commonly cited in the literature~\cite{grispos2016enhancement}. The responses are summarised in Table \ref{tab:Invest_Reqs}.

\begin{table}[h]
	\caption{Consideration of Requirements for Investigations}
	\vspace{-10pt}
	\label{tab:Invest_Reqs}
	\begin{center}
		\begin{tabular}{ | l | r | r | r | }
			\hline
			
			\multicolumn{1}{|l|}{\textbf{Requirements to assist with}} & \multicolumn{1}{|c|}{\textbf{Yes}} & \multicolumn{1}{|c|}{\textbf{No}} & \multicolumn{1}{|c|}{\textbf{DNK}} \\ \hline
			
			the detection of incidents & 63.8\% & 33\% & 3.2\%  \\  \hline
			
			the investigation of incidents & 34\% & 60.7\% & 5.3\%  \\  \hline
			
			the eradication of incidents & 28.7\% & 63.8\% & 7.5\%  \\  \hline
			
			the recovery from incidents & 36.2\% & 58.5\% & 5.3\%  \\  \hline
			
		\end{tabular}
	\end{center}
	\vspace{-10pt}
\end{table}

The table shows that nearly 64\% of the respondents indicated that their organisation consider requirements for the detection of incidents. However, a smaller number of participants stated that their organisations consider the other requirements. 34\% of the respondents indicated that requirements to assist with incident investigations were considered in their organisation, and nearly 29\% of the respondents stated that their organisation considers requirements to assist with the eradication of incidents. Moreover, approximately 36\% of the respondents stated that their organisation considered requirements to aid the recovery from incidents. 

The participants were then asked where these requirements were first discussed in the development process, with the results summarised in Table \ref{tab:Investigation_Involvement}. Note respondents were able to select more than one answer. The variety of answers that were received from this question shows that organisations appear to consider such requirements throughout the development lifecycle. However, the results also indicate that in some organisations these requirements are considered after deployment (post-delivery). Consequently, this could result in increased development costs and delayed release times \cite{chung2012non}.

\begin{table}[h]
	\caption{Requirements for Investigations -- Stages}
	\vspace{-10pt}
	\label{tab:Investigation_Involvement}
	\begin{center}
		\begin{tabular}{ | l | r |  }
			\hline
			
			\multicolumn{1}{|c|}{\textbf{Stage}} & \multicolumn{1}{|c|}{\textbf{Percentage}} \\ \hline
			
			During the problem definition/analysis phase & 46.4\%  \\  \hline
			
			During the solution development phase & 19.1\%  \\  \hline
			
			During the testing or evaluation phase & 29.8\%  \\  \hline
			
			Post-delivery & 47.6\%  \\  \hline
			
			Don't Know & 1.1\%  \\  \hline
			
			Other & 1.1\%  \\  \hline
			
		\end{tabular}
	\end{center}
		\vspace{-10pt}
\end{table}

An aggregate analysis of these two questions presents an interesting view of the results. 16 out of the 94 respondents indicated that their organisation consider all four of the requirements in Table \ref{tab:Invest_Reqs}. These 16 respondents also indicated that their organisations consider these requirements in the problem definition/analysis and solution phases. In other words, organisations that are considering all the requirements in Table \ref{tab:Invest_Reqs} are discussing them early on in the development process.  

We then asked the participants, who had indicated in the previous question that their organisation considered requirements for investigations, about the stakeholders involved in these requirements. The results are presented in Table \ref{tab:Invest_Stakes}, with the participants able to select more than one option. The table shows that a variety of stakeholders are involved in the consideration of these requirements. This includes software/system engineers (41.7\%), software/project managers (39.3\%) and security engineers (41.2\%). However, the participants indicated that other types of stakeholders are also involved in this process. For example, nearly 35\% of the respondents indicated that security incident response team members are involved in the elicitation of these requirements within their organisation. These individuals are usually responsible for conducting investigations within organisations \cite{grispos2015security}. The most common answer from the `Other' responses was `business analyst', with four respondents indicating this was the case.  

\begin{table}[h]
	\caption{Requirements for Investigations -- Stakeholders}
	\vspace{-10pt}
	\label{tab:Invest_Stakes}
	\begin{center}
		\begin{tabular}{ | l | r |  }
			\hline
			
			\multicolumn{1}{|c|}{\textbf{Stakeholder}} & \multicolumn{1}{|c|}{\textbf{Percentage}} \\ \hline
			
			Customers & 21.4\%  \\  \hline
			
			End-Users & 22.6\%  \\  \hline
			
			Software/Systems Engineers & 41.7\%  \\  \hline
			
			Software/Project Managers & 39.3\%  \\  \hline
			
			Security Analysts & 44.1\%  \\  \hline
			
			Security Engineers & 41.2\%  \\  \hline
			
			Security Managers & 41.2\%  \\  \hline
			
			Security Incident Response Team Members & 34.5\%  \\  \hline
			
			Others & 6.4\%  \\  \hline
			
		\end{tabular}
	\end{center}
	\vspace{-10pt}
\end{table}


\subsection{Requirements for Forensic Data}
\label{subsec:datareqs}

In order to examine the causes of an incident, investigators need access to forensic data \cite{carrier2003getting}. Hence, we proceeded to ask questions surrounding the consideration of requirements for the availability, collection, secure storage, tamper-proofing and examination of data for forensics. Table \ref{tab:Data_Reqs} summarises the participant's answers to these questions. 

\begin{table}[h]
	\caption{Consideration of Requirements for Forensic Data}
	\vspace{-10pt}
	\label{tab:Data_Reqs}
	\begin{center}	
		\begin{tabular}{ | l | r | r | r | }
			\hline		
			\multicolumn{1}{|l|}{\textbf{Requirements for the...}} & \multicolumn{1}{|c|}{\textbf{Yes}} & \multicolumn{1}{|c|}{\textbf{No}} & \multicolumn{1}{|c|}{\textbf{DNK}} \\ \hline
			
			availability of data for investigations & 56.4\% & 43.6\% & 0\%  \\  \hline
			
			collection of data for investigations & 23.4\% & 71.3\% & 5.3\%  \\  \hline
			
			secure storage of data for investigations & 31.9\% & 63.8\% & 4.3\%  \\ \hline
			
			tamper-proofing forensic data & 28.7\% & 63.8\% & 7.5\%  \\  \hline	
			
			examination of forensic data & 25.5\% & 69.2\% & 5.3\%  \\  \hline
			
		\end{tabular}
	\end{center}
	\vspace{-10pt}
\end{table}

The table shows that more than half of the respondents indicated that their organisation consider requirements about the availability of forensic data. However, if forensic data is to be used during an investigation, it needs to be collected and stored correctly \cite{casey2011digital}. Just over 23\% of the respondents indicated that their organisation considers requirements to assist with the collection of forensic data, while approximately 32\% of the respondents stated that their organisation considers requirements to assist with the secure storage of forensics data.

Before an investigator can collect forensic data related to an incident, malicious actors can modify this data thereby influencing the outcome of any future investigation \cite{casey2011digital}. Requirements to prevent the tampering of forensic data could be one solution to this problem. Around 29\% of respondents indicated that their organisations consider requirements to prevent the tampering of forensic data. Moreover, approximately 25\% of the respondents indicated that their organisation considers requirements to assist with the examination of data. 

We then proceeded to ask the participants, who indicated that their organisation considered these requirements about stakeholder involvement. The responses to this query are presented in Table \ref{tab:Data_Stakes}, with respondents able to select more than one option. Similar to the findings discussed in Section \ref{subsec:investigationreqs}, the respondents indicated that numerous stakeholders are involved in the elicitation of these requirements. These include software/system engineers (34.8\%), software/project managers (27\%), security managers (30.3\%), security engineers (31.5\%) and security incident response teams (22.5\%). An analysis of the `Other' (10.1\%) responses revealed that `other' stakeholders include legal and regulatory officers, software architects and governance teams. This indicates that these requirements are not just forensic goals, but legal, regulatory and governance goals as well. 

\begin{table}[h]
	\caption{Requirements for Forensic Data -- Stakeholders}
	\vspace{-10pt}
	\label{tab:Data_Stakes}
	\begin{center}
		\begin{tabular}{ | l | r |  }
			\hline
			
			\multicolumn{1}{|c|}{\textbf{Stakeholder}} & \multicolumn{1}{|c|}{\textbf{Percentage}} \\ \hline
			
			Customers & 14.6\%  \\  \hline
			
			End-Users & 11.3\%  \\  \hline
			
			Software/Systems Engineers & 34.8\%  \\  \hline
			
			Software/Project Managers & 27\%  \\  \hline
			
			Security Analysts & 36\%  \\  \hline
			
			Security Engineers & 31.5\%  \\  \hline
			
			Security Managers & 30.3\%  \\  \hline
			
			Security Incident Response Team Members & 22.5\%  \\  \hline
			
			Other & 10.1\%  \\  \hline
			
		\end{tabular}
	\end{center}
	\vspace{-10pt}
\end{table}

After identifying the associated stakeholders, we then asked the participants where these requirements were first integrated in the development process. The results are shown in Table \ref{tab:Data_Involvement}, with respondents able to select more than one answer. 

\begin{table}[h]
	\caption{Requirements for Forensic Data -- Stages}
	\vspace{-10pt}
	\label{tab:Data_Involvement}
	\begin{center}
		\begin{tabular}{ | l | r |  }
			\hline
			
			\multicolumn{1}{|c|}{\textbf{Stage}} & \multicolumn{1}{|c|}{\textbf{Percentage}} \\ \hline
			
			During the problem definition/analysis phase & 42.7\%  \\  \hline
			
			During the solution development phase & 18\%  \\  \hline
			
			During the testing or evaluation phase & 19.1\%  \\  \hline
			
			Post-delivery & 27\%  \\  \hline
			
			Don't Know & 2.2\%  \\  \hline
			
			Other & 2.2\%  \\  \hline
			
		\end{tabular}
	\end{center}
	\vspace{-8pt}
\end{table}

The responses to this question shows that organisations appear to consider such requirements throughout the development lifecycle. In summary, 42.7\% of the participants indicated that these requirements are considered during the problem definition/analysis phase, while 27\% indicated that their organisations consider such requirements post-delivery.

\subsection{Practitioner's Perception of Requirements for Forensics}

\begin{figure}[h]
	\caption{Characterising Requirements for Forensics}
	\label{fig:Characteristics}		
	\centering
	
	\includegraphics[trim=50 80 100 90, scale=0.33]{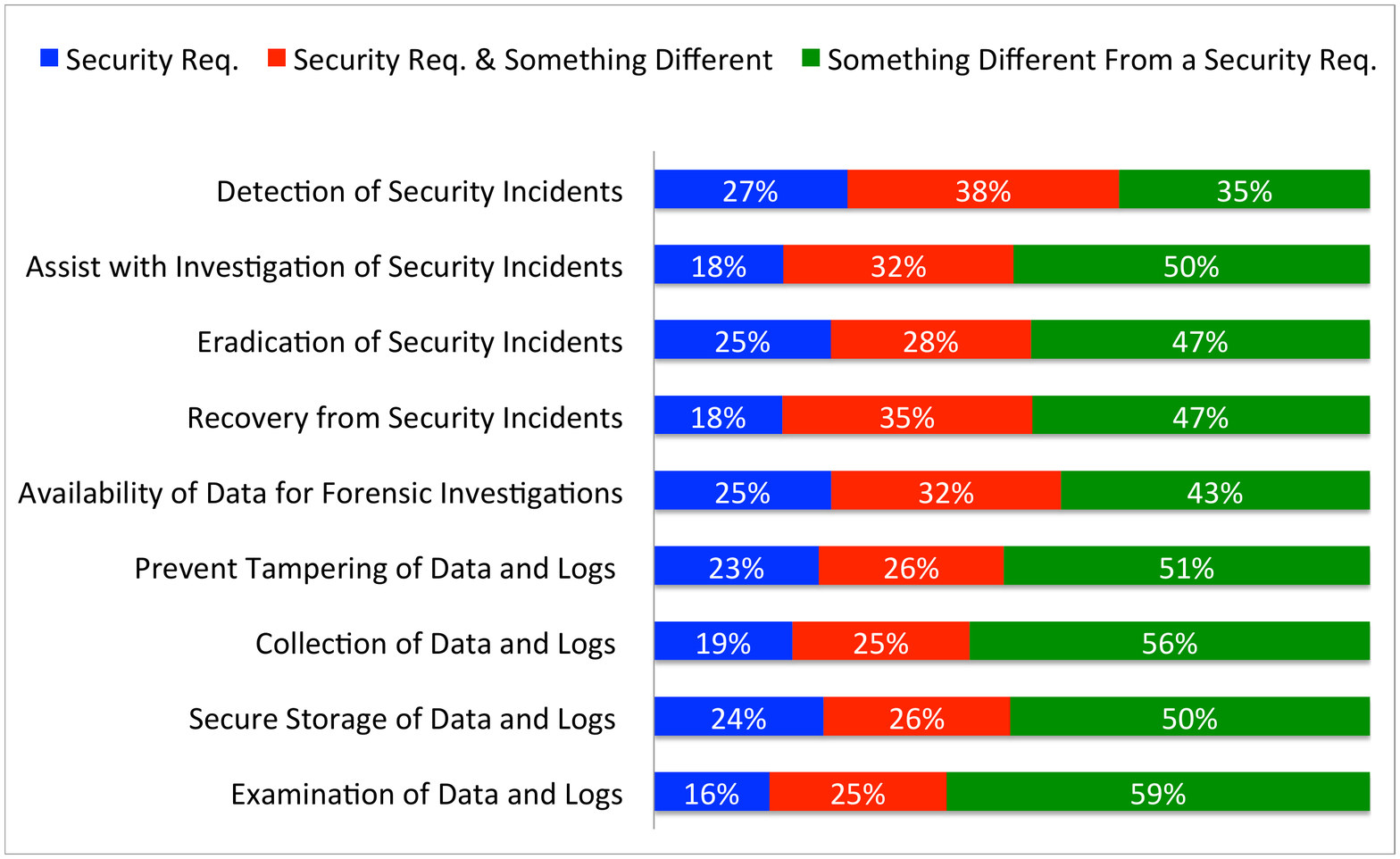}
	\vspace{-8.5pt}
	
\end{figure}

In order to examine how practitioners characterise requirements for forensics, we then asked the participants to rate the requirements presented in Table \ref{tab:Invest_Reqs} and Table \ref{tab:Data_Reqs} on a 3-point scale of: ``Security Requirement'', ``Mixture of Security Requirement and Something Different'' and ``Something Different from a Security Requirement''. The results are presented in Figure \ref{fig:Characteristics} with the values presented rounded to the nearest percentage. The figure shows that for eight out of the nine requirements, the most common answer was ``Something Different from a Security Requirement''. With regard to the remaining requirement (Detection of Security Incidents), the most common answer provided was a ``Mixture of Security Requirement and Something Different''.

In addition to the above queries, the respondents were provided with a definition of digital forensics and asked ``does your organisation elicit requirements for digital forensics during its development lifecycle?''. The purpose of this question was to evaluate the respondent's awareness surrounding requirements for forensics based on the previous questions. In response, 23.4\% stated that their organisation \textit{does} elicit requirements for forensics. However, 71.3\% indicated that their organisation \textit{does not} elicit such requirements, while 5.3\% provided `do not know' answers. Examining the `No' answers with the answers provided to questions described in Sections \ref{subsec:investigationreqs} and \ref{subsec:datareqs} presents an interesting observation. 49 of the respondents (73\%) who provided `No' answers to the above question, indicated in previous questions that their organisation \textit{did} consider requirements for investigations and forensic data. These findings could indicate limited digital forensics awareness among the participants. 


\subsection{Threats to Validity}

Our empirical study involved 94 respondents in order to assess the extent to which requirements for forensics are considered within organisations and to identify research challenges that emerge from this work. However, with research of this nature there is a need to evaluate the threats to validity, which we analysed using a scheme presented by Runeson, et al. \cite{Run2012CaseStudy}. From a \textit{construct validity} perspective, it must be acknowledged that there is a possibility that the respondents did not fully understand the questions or chose to respond arbitrarily. To mitigate this threat, the survey was validated by three academic researchers and two software engineering professionals. The researchers were within the domains of requirements engineering, security and digital forensics, while the two professionals are employed in organisations that are involved in software development. The validation was done by taking the survey and providing feedback, which was then used to improve question phrasing and adding answer options. In addition, we included validation questions in the survey and have only reported responses from the respondents who correctly answered those questions. 

In terms of \textit{internal validity}, it must be acknowledged that we relied on LinkedIn to recruit survey participants. Hence, the results obtained in the survey may reflect only the opinion of the individuals on this social network within the specific groups that were targeted. This threat is partly alleviated due to the selection of groups on LinkedIn that focused on requirements and software engineering, information security and digital forensics. A potential threat to \textit{external validity}, is that we only obtained 94 answers from practitioners. Nevertheless, these individual’s experience, industries and job roles are widespread and we are confident that our results present a broad representation of requirements for forensics practices within organisations. With regard to \textit{reliability}, the survey data was analysed independently, first by the primary author and then the secondary author. This also provided consistency regarding the analysis of the data.

\section{Research Challenges}
\label{sec:Challenges}

Our findings, coupled with the results from the literature review has prompted the identification of several research challenges, which we discuss below. 

\subsubsection{\textbf{Elicitation of Requirements for Forensics}}

Our findings show that organisations consider requirements for forensics during various stages in the development process, including post-delivery. As software engineering researchers and practitioners \cite{chung2012non} have discussed, non-functional requirements (e.g. requirements for forensics) can impact a software system's architecture. Hence, we propose that organisations consider requirements for forensics at the start of development process. However, techniques for eliciting and analysing these requirements have not actually been proposed in the literature. Therefore, there is also a need to examine if existing elicitation techniques can be used for `forensic requirements'.  
 
\subsubsection{\textbf{Stakeholder Analysis for `Forensic Requirements'}} Our analysis of the literature and survey findings have indicated that a variety of stakeholders are likely to be involved in the consideration of requirements for forensics. Previous work~\cite{sharp1999stakeholder} has argued that the effective consultation of relevant stakeholders is very important in the requirements engineering process. We envision that the same would apply to stakeholders involved in `forensic requirements'. Hence, there is a need to investigate how these differing interests and demands in requirements for forensics can be accommodated from these multidisciplinary stakeholders.


\subsubsection{\textbf{Evaluating the Impact of Laws and Regulations on Requirements for Forensics}}

Previous work \cite{garcia2016towards} has argued that different laws and regulations can impact the elicitation of compliance requirements. However, further research is needed to examine how these conflicts could influence requirements for forensics. For example, the Cybersecurity Information Sharing Act in the United States could conflict with European data protection regulations. Hence elicitation approaches for `forensic requirements' could also need to take into account laws and regulations from multiple jurisdictions.

\subsubsection{\textbf{Forensics Trade-off Analysis}}

The differing viewpoints from which requirements for forensics can be elicited could inevitably cause conflicts. As a result, engineering forensic-ready systems is likely to require trading off `forensic requirements' with other requirements, such as security and privacy. Therefore, there is a need to first, examine the applicability of existing trade-off techniques \cite{Karlsson1996Software} to support forensics trade-off analysis and second, if required, develop new techniques to support trade-off analysis between `forensic requirements' and other requirements.

\subsubsection{\textbf{Addressing System Performance Overhead in Forensic-by-Design}}

Multiple authors have discussed the concept of `system forensic readiness' as a means of implementing forensic readiness \cite{tan2001forensic,barske2010digital,pooe2012conceptual,rowlingson2004ten,yasinsac2001policies}. Inline with the concept of `system forensic readiness', we envision that system performance challenges could emerge when implementing requirements for forensics. For example, one specific requirement could involve intensified logging of user activity, which could result in increased overhead and reduced system performance. Hence, existing research needs to be extended to analyse the system performance challenges that could emerge due to the implementation of `forensic requirements'.

\subsubsection{\textbf{Assessing the Influence of Forensic-by-Design on the Forensic Readiness Ecosystem}}

We define a \textit{forensic readiness ecosystem} as the different elements and approaches that an organisation can implement to become forensic-ready. This can include implementing policies and processes, aligning technology with forensic aims and training employees. However, existing research has overlooked the impact of forensic-by-design on this forensic readiness ecosystem. For example, an organisation may need to enhance its policies and procedures based on the requirements it will consider during the engineering of forensic-ready systems. Hence, empirical studies are needed to assess the impact of forensic-by-design on the wider forensic readiness ecosystem.

\section{Conclusions}
\label{sec:Conclusions}

In this paper, we undertook an online survey targeting industry practitioners to assess and understand the extent to which requirements for forensics are recognised explicitly and implemented within organisations. The survey results suggest that although organisations appear to consider such requirements, a number of inconsistencies emerge. These inconsistencies include when such requirements are actually introduced in the development process, as well as the stakeholders involved in their elicitation. Our analysis, coupled with a literature review on forensic readiness, led us to recognise a number of research challenges for the explicit elicitation, analysis and implementation of such requirements. Future work intends to further examine and validate our findings. This will be done through in-depth interviews with practitioners identified in this survey as well as empirical case studies to further investigate organisational forensic-by-design practices. Future work also intends to explore the possibility of using existing approaches that are currently used to elicit security and legal requirements in a `forensic requirements' context. Any proposed solution will also need to take into account the integration of `forensic requirements' into agile software development, techniques to identify conflicts between requirements for forensics with other requirements and support trade-off analysis.

\section*{Acknowledgments}

This work was partially supported by SFI Grants No. 13/RC/2094 and 15/SIRG/3501 and ERC Advanced Grant. No. 291652 (ASAP).

\bibliographystyle{IEEEtran}
\bibliography{RCIS}

\end{document}